\documentstyle[aps,preprint]{revtex}
\begin{document} \draft

\title {Relaxation in a perfect funnel}

\author{Maxim Skorobogatyy, Hong Guo and Martin Zuckermann}

\address{Department of Physics and\\
Centre for the Physics of Materials\\
McGill University\\
Montr\'eal, Qu\'ebec, H3A 2T8 Canada.}

\date{\today}
\maketitle

\begin{abstract}

We have exactly solved the relaxational dynamics of a model protein
which possesses a kinetically perfect funnel-like energy landscape.
We find that the dependence of the relaxation time, $\tau$, on the density 
of states (DOS) and the energy level spacing distributions of the model
displays several main types of behavior depending on the temperature $T$.
This allows us to identify possible generic features of the relaxation. 
For some ranges of $T$, $\tau$ is insensitive to the density of states;
for intermediate values of $T$ it depends on the energy level spacing 
distribution rather than on the DOS directly, and it becomes gradually 
more dependent on DOS with increasing temperature; finally,
the relaxation can also be determined exclusively by the presence
of a deep gap in the energy spectrum rather than by the detailed 
features of the density of states. We found that the behavior of 
$\tau$ crucially depends on the degeneracy of the energy spectrum.  For
the special case of exponentially increasing degeneracy, we were able to 
identify a characteristic temperature which roughly separates the 
relaxational regimes controlled by energetics and by entropy, 
respectively. Finally, the validity of our theory is discussed when 
roughness of energy landscape is added.  

\end{abstract}

\vspace{0.5in}


\baselineskip 16pt

\newpage

\section{Introduction}

It is well known that natural proteins fold into their native structures
remarkably quickly in times on the order
of a second in spite of the enormous number of possible physical 
configurations~\cite{creighton}. On the other hand, it is also clear that
heteropolymers with completely random monomer-monomer interactions usually
do not fold on a reasonable time scale\cite{bryngelson}.  One explanation
put forward to resolve this discrepancy is that protein sequences are 
``optimized'' such that not only is there a stable unique 
structure for the ground state \cite{go}, but there is also a funnel-like 
energy landscape which leads to efficient folding kinetics 
\cite{montal,wolynes,onuchic}. A principle of minimal frustration was
proposed\cite{bryngelson1} to enforce a selection of the interactions 
between monomers such that as few energetic conflicts occur as possible.
Among other things, considerable theoretical effort has concentrated on finding 
proper models for protein folding and investigating various sequencings
which lead to fast folding kinetics. Due to the immense complexity 
of the problem, much of our understanding and intuition has been obtained
from a variety of computer simulations based on lattice 
models\cite{honeycutt,shakhnovich2}.

In this paper we concentrate on the folding scenario involving
a funnel-like energy landscape\cite{montal,wolynes,onuchic} 
where the funnel ``guides'' the protein into
the low energy native structure. Along the pathway, the protein is
believed to go through several distinct states including the molten globule
state, a folding transition region, and a glass transition region. Even though
the funnel landscape possesses a certain amount of roughness which slows 
the folding kinetics, the folding process is largely speaking guided by 
the global funnel structure and the protein can in this way rapidly find 
its native state. Although there is no clear experimental evidence of 
the existence of this folding scenario, it is nevertheless theoretically 
interesting and attracted much attention in the literature.  In general, the 
folding kinetics for funnel--like energy landscapes is very complicated and 
analytical studies have proved to be quite difficult. In this regard, 
an interesting study is due to Zwanzig\cite{zwanzig} where some of the 
general properties of the folding kinetics were examined using an extremely 
simple model. 

Even though there are many folding models for specific
proteins, we believe that it is intuitively useful to
investigate the generic behavior of the folding kinetics. This, 
in some sense, is similar to finding universality classes in 
critical phenomena. The purpose of this paper is to report our studies 
in this direction. In particular we examine a simple statistical mechanical
model which mimics all the basic properties of a perfect funnel-like landscape 
in the absence of roughness. We also discuss the validity of our results 
for the case of involving small amount of roughness (see below). The landscape 
itself consists of a set of energy levels forming a quasi-continuous 
spectrum with a single level lying far below this spectrum. All the levels 
represent conformational energies of the protein with the lowest level 
representing the native state.  This model is quite general and is not 
exclusive to proteins. It could for example represent certain classes 
of polymers.

There are many interesting questions concerning protein folding kinetics
which we would like to answer from an analysis of our model. For
instance, for a protein sequence which folds rapidly to its 
native state, what is the role played by the energy spectrum along 
the folding pathway ? What is the role of energy level spacing statistics 
on the folding kinetics ?  How does the relaxation process of our system
depend on parameters such as temperature ? These are interesting 
and difficult general questions which are relevant to the folding 
kinetics, since the protein passes through the energy spectrum during the 
folding process. An analytical answer to these questions for a general 
protein problem has so far been not possible. However as we show below for 
our model which is a generalization of that studied in Ref. \cite{zwanzig}, 
analytical solutions can be found when the funnel structure has no roughness
and reasonable approximations could be made to find an answer when small
roughness is included. Our perspective is that exact solutions are valuable since 
they can be used as a starting point for further more complicated 
models, similar to our experiences in critical phenomena and phase transitions. 

To this purpose, we have derived analytical expressions for our model 
(see below) which show that the energy spectrum and the energy spacing 
statistics can play an important role in folding kinetics. In particular 
the dependence of the folding time, $\tau$, on the energy level 
distribution of the various models can be classified into three main 
types of behavior depending on the temperature $T$. For a considerable 
range of $T$, $\tau$ is insensitive to the level distribution; 
for intermediate values of $T$ it depends directly on the distribution and 
becomes gradually more model dependent;  finally, in the third case 
the folding kinetics is determined exclusively by the presence of the 
deep gap in the energy spectrum rather than by the details of the energy 
level distribution. We found that the behavior of $\tau$ crucially depends 
on the degeneracy of the energy spectrum.  For the special case of 
exponentially increasing degeneracy, we were able to identify a 
characteristic temperature which roughly separates the relaxational 
regimes controlled by energetics and by entropy respectively.  Our general 
formula for this simple model is consistent with existing literature in 
the appropriate limits and we present numerical solutions to confirm the
physical picture indicated by our analytical results.

The paper is organized in the following manner.  A general expression for
the relaxation time is derived in the next section. Sections III 
and IV presents results for the applications of this expression. 
Finally a short summary is presented in the last section.

\section{Relaxation Kinetics}

By analogy with Ref.~\onlinecite{zwanzig}, we focus on a perfect 
funnel-like energy landscape defined by an abstract ``reaction 
coordinate'' $X$.  For example $X$ could represent a specific 
protein structure which has energy $E(X)$. We emphasis again that 
whether or not a real protein possesses funnel-like energy landscape 
is unclear, but we shall examine the consequences of
this landscape. A perfect funnel with no roughness is schematically 
shown in Fig. (\ref{funnel}). Clearly this is a considerable 
simplification of the problem, but it allows us to investigate the 
relaxational kinetics completely analytically. As discussed below, 
other features can be systematically added on later, such as a small 
amount of roughness.

Thus, as the system relaxes or the ``protein'' folds, it rolls down
the funnel, $E(X)$, to the final native structure characterized by energy
$E_o$. Now a particular model can be described in terms of 
its energy level distribution or its density of states. 
Here we consider the situation where there are $N$ 
quasi-continuous energy levels with density of states, $D(E)$, and one 
distant level, $E_o$, lying at a distance $\Delta E_{o}$ below the 
quasi-continuous spectrum (see Fig. (\ref{funnel})). During the folding 
process (relaxation process) we assume~\cite{zwanzig} that in any transition 
between configurations, $X$, changes only by $\pm 1$ which means that the 
system performs a nearest neighbor random walk in one-dimensional reaction 
coordinate space. For the perfect funnel energy landscape considered here,
$X \rightarrow X \pm 1$ also implies that energy transitions only 
occur between nearest 
neighboring levels. In this work we use Metropolis transition rates, same as
that used in Ref. \cite{zwanzig}, which is justified since no roughness 
is included in the model. On the other hand if the landscape is very 
rough indicating entanglement of the polymer or protein, Metropolis rates 
will not be adequate. The Metropolis rates satisfies
the requirement of detailed balance
\begin{equation}
W(X\rightarrow X+1)P_{X}(eq)=W(X+1\rightarrow X)P_{X+1}(eq) \ \ .
\end{equation}  
where $P_{X}(eq)$ is the equilibrium distribution given by
a Boltzmann factor.
Using $g_{X}$ and $g_{X+1}$ as the degeneracies of the  
$X$ and $X+1$ energy levels, one can then introduce the following
transition rates
$W(X\rightarrow X+1)=\frac{g_{X+1}}{g_{X}}\exp{-(E_{X+1}-E_{X})
\over{T}}$ and $W(X+1\rightarrow X)=1$. The second condition, which is
independent of temperature $T$, corresponds to the zero-roughness on the 
$E(X)$ landscape.

The folding or relaxation kinetics is studied using a master equation.  
We focus on the probability, $P_i$ ($i=0,1,2,...N$), of being at energy 
level $E_i$ during the relaxation. Introducing variables $\alpha_{i} \equiv
\exp{-\frac{\bigtriangleup F_{i}}{T}}$ where
$\bigtriangleup F_{i}=F_{i+1}-F_{i} \ i\in\{0,N-1\}$, and 
$F_{i}=E_{i}-T\ln{g_{i}}$
we can write down the following matrix equation for the 
evolution of the probabilities
\begin{equation}
\frac{d\bar{P}}{dt}=A\bar{P}\ \ ,
\label{eqn1}
\end{equation}
where the matrix coefficient is given by
\begin{equation}
A = \left\| \begin{array}{ccccc}
-\alpha_{0}  &1             &0             &0             &\cdot \\ 
\alpha_{0}   &-1-\alpha_{1} &1             &0             &\cdot \\
0            &\alpha_{1}    &-1-\alpha_{2} &1             &\cdot \\
0            &0             &\alpha_{2}    &-1-\alpha_{3} &\cdot \\
\cdot &\cdot &\cdot &\cdot &\cdot
\end{array}   
\right\|
\end{equation}

Because the total probability is a constant, the matrix $A$ is thus 
degenerate and there are only $N$ independent probabilities out of a total 
of $N+1$. We denote the average nearest neighbor energy level spacing 
by $\bar{U}$. We then make the reasonable 
assumption that $\bar{U} \ll \bigtriangleup E_{o}$ which is basically a 
consequence of having a 
spectrum with a few low lying energy levels and a quasicontinuous part 
in the upper part of a spectrum. Several limits can be obtained directly from 
the form of the matrix $A$. 
For $0\leq T\ll \bar{U}$ and $\bar{U}\ll T\ll \Delta E_o$, 
$A$ essentially becomes a constant matrix largely independent of 
temperature except with only a few temperature dependent terms
and thus the folding kinetics is independent of the energy spectrum, $D(E)$. 
The same is true for very high temperatures where entropy is the dominate 
factor. At temperatures $T \sim \Delta E_o \gg \bar{U}$, almost all matrix 
elements of $A$ become constants except those few involving energy differences 
comparable to the $\Delta E_{o}$. This suggests that in this case the 
folding kinetics is 
exclusively determined by the presence of the gap in the energy spectrum 
while almost all quasi-continuous levels are already excited.

A non-trivial result was obtained for the temperature range 
$T \sim \bar{U}$. Here the kinetics can substantially depend on the energy 
distribution of the quasi-continuous part of the spectrum.
We first find a relaxation time at temperatures small enough 
that one can disregard the rate of escape from the native state. 
In this range the master equation of (\ref{eqn1}) becomes 
$dP_o/dt = P_1$ and 
\begin{equation}
\frac{d\bar{P'}}{dt}\ =\ M\cdot \bar{P'}\ .
\label{eqn2}
\end{equation}
where $\bar{P'}$ is a vector $(P_{1},P_{2},...,P_{N})$.
The matrix $M$ is a sub-matrix of $A$ without its first row and first column.
As the relaxation proceeds, {\it i.e.} when our system rolls down the perfect 
funnel $E(X)$, the total relaxation time from the highest energy level
$E_N$ to the lowest one $E_o$, gives a measure of the relaxation time. 
To determine this relaxation time, $t_{rel}$, we notice that the system 
has relaxed  when all states with $P_{i}$, $i\in\{1,N\}$ have been 
sequentially relaxed. 
Therefore $t_{rel} \sim \sum_{i=1}^{N} t_{i}$ where $t_{i}$ are the
relaxation times for the states with $P_{i}$.  

It is worth pointing out that at low temperatures the relaxation time
of a system coincides with its folding time into the ground (native) state
because the native state at these temperatures is an equilibrium state 
of the system. As temperature increases, however, the equilibrium state 
of the system shifts to the quasicontinuous part of the spectrum.
Hence $t_{rel}$ will characterize the folding time to the appropriate 
equilibrium. Keeping this in mind, we now calculate $t_{rel}$.
From Eq. (\ref{eqn2}) we
have $\sum_{i=1}^{N} t_{i} = -\sum_{i=1}^{N} \frac{1}{\lambda_{i}}$ where 
$\lambda_{i}$ are the eigenvalues of the matrix $M$. As $det(M)\neq 0$,
$1\over{\lambda_{i}}$ are the eigenvalues of matrix $M^{-1}$. In this way 
we finally obtain $t_{rel}$ in terms of the trace of $M^{-1}$ since
$t_{rel}\sim -Tr(M^{-1})$.

The calculation of $Tr(M^{-1})$ is quite lengthy~\cite{maxim1} and we
outline the main steps here. We notice that $M=L+\delta L$ where $L$
is a constant matrix and $\delta L$ contains the temperature dependent
elements of $M$. We seek $M^{-1}$ in a perturbative form given by 
$M^{-1}=\sum_{i=0}^{+\infty} S^{(i)}$. It turns out that this sum 
only has a finite number of non-vanishing terms and can thus be summed
exactly. Essentially, from 
$M\cdot M^{-1}=I$ one derives a set of equations for $S^{(i)}$
which can be solved to give $S^{(i)}=(-1)^{i}(S^{0}\delta L)^{i}S^{0}$,
where $S^{0}$ is a triangular matrix with all the upper right 
elements equal to $-1$. Because of the simplicity of $\delta L$ and $S^{0}$,
$(S^{0}\delta L)^{i} = 0$ for all $i \geq N$. 
After lengthy but straightforward algebra we obtain 
\begin{equation}
M^{-1} = \left\| \begin{array}{ccccc}
1&0                    &0             &0             &\cdot \\
\alpha_{1} &1                    &0             &0             &\cdot \\
\alpha_{1}\alpha_{2}  &\alpha_{2} &1             &0             &\cdot \\
\alpha_{1}\alpha_{2}\alpha_{3} &\alpha_{2}\alpha_{3} &\alpha_{3} &1 &\cdot \\
\cdot &\cdot &\cdot &\cdot &\cdot
\end{array}
\right\|\cdot S^{0}\ \ \ .
\label{mm1}
\end{equation}
\begin{equation}
\tau=-Tr(M^{-1})=N+\alpha_{1}+\alpha_{2}+\cdot \cdot \cdot +\\ 
\alpha_{1}\alpha_{2}+\alpha_{2}\alpha_{3}+\cdot \cdot \cdot +\\
\alpha_{1}\alpha_{2}\alpha_{3}+\cdot \cdot \cdot\\
\end{equation}

We now solve the problem in general for all T.
To find the relaxation time $\tau$ of this perfect-funnel model, we use 
the fact that the total probability is conserved, i.e. that
\begin{equation}
P_{0}=1-\sum_{i=1}^{N} P_{i}\ \ .
\end{equation}
Then equation (2) is equivalent to the following equation,
\begin{equation}
\frac{d\bar{P'}}{dt}\ = 
\alpha_{0}\delta_{i,0}+(M-\alpha_{0}\delta M)\ 
\cdot \bar{P'}\ \ .   
\end{equation}
Thus the kinetics will basically be determined by the 
matrix $M-\alpha_{0}\delta M$ where $\delta M$ is a matrix 
with a first row consisted of $1$ and the rest of the elements 
equal zero. As discussed above, the relaxation time of the 
system can be found as the trace of $(M-\alpha_{0}\delta M)^{-1}$. 
To compute this trace we use the perturbation expansion 
introduced above and seek an inverse matrix in the following form 
\begin{equation}
(M-\alpha_{0}\delta M)^{-1}=H_{(0)}+H_{(1)}+H_{(2)}+\cdot\cdot\cdot\ \ \ .
\label{inverse}
\end{equation}
Using the identity 
\[
I\equiv (M-\alpha_{0}\delta M)(H_{(0)}+H_{(1)}+H_{(2)}+\cdot\cdot\cdot)
\]
we find a set of equations for $H^{(i)}$ which can be solved to 
give $H^{(i)}=M^{-1}(\alpha_{0} \delta MM^{-1})^{i}$, where $M^{-1}$ is found 
earlier and is given by Eq. (\ref{mm1}).
Hence,
\[
(M-\alpha_{0}\delta M)^{-1}=M^{-1}(1+\alpha_{0}(\delta MM^{-1})
+\alpha_{0}^{2}(\delta MM^{-1})^{2}+\cdot\cdot\cdot\ \ \ .
\]
Introducing a new quality 
\begin{equation}
R_{i}\equiv 1+\alpha_{i}+\alpha_{i}\alpha_{i+1}+
\alpha_{i}\alpha_{i+1}\alpha_{i+3}+\cdot\cdot\cdot+
\alpha_{i}\cdot\cdot\cdot\alpha_{N}
\end{equation} 
one can easily show that the required inverse matrix can be written as
\begin{equation}
(M-\alpha_{0}\delta M)^{-1}=M^{-1}+
\alpha_{0}M^{-1}\delta MM^{-1}(1-(\alpha_{0}R_{1})+
(\alpha_{0}R_{1})^{2}-(\alpha_{0}R_{1})^{3}+\cdot\cdot\cdot\ \ .
\label{mm2}
\end{equation}
This is an important result and the trace of this matrix gives the relaxation
time of our model.

{\it Formally} the sum in the bracket of Eq. (\ref{mm2}) 
is a geometric series, which we can rewrite in a more compact form,
\begin{equation}
(M-\alpha_{0}\delta M)^{-1}=M^{-1}+
\frac{\alpha_{0}}{1+\alpha_{0}R_{1}}M^{-1}\delta MM^{-1}\ \ .
\label{mm2a}
\end{equation}
Strictly speaking the summation of the series is allowed only when
$(\alpha_{0}R_{1})<1$. Thus in principle we should use Eq. (\ref{mm2}) when
this condition is not satisfied. It turns out, by numerical comparisons,
that Eq. (\ref{mm2a}) is correct even for $(\alpha_{0}R_{1})>1$.
Even though this indicates that there is most probably a more direct 
way of deriving Eq. (\ref{mm2a}) rather than via the series expansion 
used here, we use (\ref{mm2a}) to proceed further and present
numerical confirmation of this procedure later.

The relaxation time is then given by
\begin{equation}
\tau=-Tr(M^{-1})-\frac{\alpha_{0}}{1+\alpha_{0}R_{1}}
Tr(M^{-1}\delta MM^{-1})\ \ .
\end{equation}
Using the explicit forms of the matrices as given by Eq. (\ref{mm1}) and
$\delta M$, after lengthy but straightforward algebra we obtain,
\begin{equation}
Tr(M^{-1}) = N + \sum_{i=1}^{N-1} \sum_{j=1}^{N-i}
\alpha_{j}\cdot\cdot\alpha_{j+i-1}\ \ \ ,
\end{equation}
and 
\[Tr(M^{-1}\delta MM^{-1})=R_{1}+(R_{1}+R_{2})\alpha_{1}+
(R_{1}+R_{2}+R_{3})\alpha_{1}\alpha_{2}+\cdot\cdot\cdot \]
\begin{equation}
+ (R_{1}+\cdot\cdot\cdot+R_{N})\alpha_{1}\cdot\cdot\alpha_{N-1}\ \ .
\end{equation}

If we define $Z_{i}$ as $Z_{i}\equiv\sum_{j=i}^{N} \exp{-\frac{F_{j}}{T}}$,
$Z_{i}$ is then the partition function for the 
$N-i$ energy levels starting at $i$ 
and $R_{i}$ is given by $R_{i}=Z_{i}\exp{\frac{F_{i}}{T}}$.
The above results can then be considerably simplified and 
the expression for $\tau$ becomes
\begin{equation}
\tau=-Tr(M^{-1})-
\frac{\sum_{i=1}^{N} 
Z_{i}^{2}\exp{\frac{F_{i}}{T}}}{Z_{0}}\ \ . 
\end{equation}

Using Eqs.~(14) to (16) we arrive at the main result of 
this work, 
\begin{equation}
\tau=\eta_1 + \eta_2
\label{result}
\end{equation}
where
\begin{equation}
\eta_1\ =\ (N+\sum_{i=1}^{N-1} 
\sum_{j=1}^{N-i}\alpha_{j}\cdot\cdot\alpha_{j+i-1})
\label{eta1}
\end{equation}
\begin{equation}
\eta_2\ =\ 
-\frac{\sum_{i=1}^{N}
Z_{i}^{2}\exp{\frac{F_{i}}{T}}}{Z_{0}}\ \ .
\label{eta2}
\end{equation}
Because this result is quite complicated, we first
apply it in the next section to various specific situations, 
and then present numerical results obtained using the general expression
of Eqs.~(\ref{result}) to (\ref{eta2}).

\section{The role of level statistics}

As a first application of the result given by Eq. (\ref{result}), we 
examined the case where the energy levels are non-degenerate, {\it i.e.} 
$g_{X}=1$ for all $X$, then ~local free energy $F_{i}$ coincides with 
the energy $E_{i}$. In this limit the expression for 
the relaxation time can be greatly simplified. Also, we shall focus on 
low temperatures.
In this case we shall prove that the relaxation time $\tau$ is determined by
the energy level spacing distributions. These distributions can be computed
from the density of states $D(E)$. Since very different models, {\it i.e.}
different $D(E)$'s, can give quite similar spacing distributions, the
relaxation in this temperature regime is quite generic.  In the next section
we shall show that even when $g_X > 1$, similar conclusions can be reached
if temperature is lower than some characteristic temperature.

\subsection{The relaxation time}

For temperatures much smaller than the energy gap between the ground
state and the first excited state, $\bigtriangleup E_o$, and taking into 
account that there are no exponentially divergent pieces in the second 
term $\eta_2$ of Eq. (\ref{result}), we can safely neglect $\eta_2$ since 
it is much smaller than the $\eta_1$ term.  Furthermore we notice that
\begin{equation}
\begin{array}{ll}
\alpha_{j}\alpha_{j+1}..\alpha_{j+i-1} &
= \exp{ -\frac{(E_{j+1}-E_{j}+E_{j+2}-E_{j+1}+
 ... + E_{j+i}-E_{j+i-1})}{T}}\\
& = \exp{-\frac{(E_{j+i}-E_{j})}{T}}
\end{array}
\end{equation}
The relaxation time $\tau$ can therefore be expressed as follows
in terms of the level spacing probability 
\[p_{(i)}(S)\equiv \sum_{j=1}^{N-i} \delta(E_{j+i}-E_{j}-S)\]
which measures the probability 
that $E_{j+i}-E_j$ equals $S$ for all level indices $j$. Using this
definition we find:
\[\sum_{j=1}^{N-i} \alpha_{j}\cdot\cdot\alpha_{j+i-1}
=\int_{0}^{+\infty}p_{(i)}(S)\exp{-\frac{S}{T}}dS\ \ . \]
From Eqs. (\ref{result}) and (\ref{eta1}) we obtain
\begin{equation}
\tau=t_{rel} \approx N + \sum_{i=1}^{N-1}
\int_{0}^{+\infty} p_{(i)}(S)\exp{-\frac{S}{T}}dS\ \ .
\label{eqn3}
\end{equation}

It is clear that small values of the level spacing, $S$, give the 
largest contribution to the integrals in (\ref{eqn3}) and this is 
especially true for low temperatures. On the other hand for higher 
order energy level spacing distributions, $p_{(i)}(S)$ is small 
for small values of $S$. Hence, for low $T$ the term
containing the nearest neighbor spacing distribution, $p_{(1)}(S)$, is most
important in determining the relaxation time $\tau$. As $T$ increases
$\tau$ begins to depend on higher order energy level spacing distributions.
Since more and more $p_{(i)}(S)$ begin to play a role as $T$ increases, 
we expect the model details to become increasingly important. However 
it is easy to show that $p_{(i)}(S)$ for large $i$ must be universal: 
as the levels are far apart there is little level correlation and the 
spacing distribution therefore approaches a Gaussian.
Hence we expect model independence to return when $T$ reaches   
values $\gtrsim \bar{U}$.  Finally, if $p_{(i)}(S)$ is less
sensitive to model peculiarities than the density of states, $D(E)$, 
we expect that the relaxation kinetics measured by the relaxation time
$\tau$ is approximately generic as it only depends on 
several low order spacing distributions in this
low temperature range. We shall confirm this picture by computing
$p_{(i)}(S)$ and in particular $p_{(1)}(S)$ in terms of $D(E)$.

\subsection{The spacing distribution}

To relate $P_{(i)}(S)$ to the density of states, $D(E)$, which specifies our
models, let us take any two energy levels and consider the probability 
that the first level lies in the interval $[E,E+dE]$ while the second 
level lies in $[E+S,E+S+dS]$. To find $p_{(i)}(S)$ we use an approximate 
approach\cite{mehta,pandey} analogous to mean field theory in which the 
energy distribution, $D(E)$, is assumed to be locally random. This 
allows us to use a simple approach based on probability theory. First we 
note that there are $i-1$ levels in the interval $[E,E+S]$ and the 
remaining $N-i$ levels are outside this interval. The 
probability for this to occur is proportional to\cite{maxim1} 
\begin{equation}
D(E)D(E+S)\left[1-\int_{E}^{E+S} D(t)dt\right]^{(N-i)}
\left[\int_{E}^{E+S} D(t)dt\right]^{(i-1)}dSdE\ \ .
\label{eqn4}
\end{equation}
Integrating over $E$ and normalizing the resulting expression, we obtain
\begin{equation}
p_{(i)}(S)=C\int_{-\infty}^{+\infty} 
D(E)D(E+S)\left[1-\int_{E}^{E+S} D(t)dt\right]^{(N-i)}
\left[\int_{E}^{E+S} D(t)dt\right]^{(i-1)}dE 
\label{eqn5}
\end{equation}
where $C$ is the normalization factor.
In spite of its involved appearance, this equation is easy to 
investigate. For example consider $p_{(1)}(S)$, which is written
\begin{equation}
p_{(1)}(S)=N(N+1)\int_{-\infty}^{+\infty} 
D(E)D(E+S)\left[1-\int_{E}^{E+S} D(t)dt\right]^{(N-1)}dE\ \ .
\label{eqn6}
\end{equation}
It is easy to see that if $D(E)$ is substantially larger than zero
on an interval $\xi$, then due to the factor 
$\left[1-\int_{E}^{E+S} D(t)dt\right]^{(N-1)}$, $p_{(1)}(S)$ will 
also be substantially larger than zero on a scale of $\frac{\xi}{N}$. 
If $S\ll 1$, in the $N\rightarrow \infty$ limit
we can expand the integrand to obtain
\begin{equation}
p_{(1)}(S)=N(N+1)\sum_{i=0}^{+\infty} 
(-1)^{i}\frac{(SN)^i}{i!}
\int_{-\infty}^{+\infty} D^{i+2}(E)dE\ \ ,
\label{eqn7}
\end{equation}
From Eq.(\ref{eqn7}) it can also be deduced that the scale on which 
$p_{(i)}(S)$ is substantially greater than zero is $O(\frac{1}{N})$. 
Furthermore, it is clear from Eqs.~(\ref{eqn5}) and~(\ref{eqn7}) 
that $P_{(i)}(S)$ 
is determined by the values of a set of definite {\it integrals} 
of the density of states distribution, $D(E)$, and its powers rather 
than the specific details of the level distribution itself.

How sensitively does $p_{(1)}(S)$ depend on $D(E)$ ?  Consider two completely
different models specified by $D_1(E)= exp(-E)$ with $E\in[0,+\infty ]$, and
$D_2(E)=\frac{d+1}{E_{0}^{d+1}}E^{d}$ with $E\in[0,E_{0}]$ and $d>0$. 
Using Eq. (\ref{eqn7}) we can explicitly compute the nearest neighbor
spacing distribution $p_{(1)}(S)$ for the two models.  It is easy to
show that both models give the same form of $p_{(1)}(S)$ for large 
values of the parameter $d$. Furthermore, even for $d\sim O(1)$ the 
difference is only $\sim (d+1)/d$. The fact that the two chosen forms 
of $D(E)$ are considerably different from one another shows that
different models described by different forms of $D(E)$ can have similar 
nearest energy level spacing distributions. Hence they can have similar
relaxation times as specified by Eq. (\ref{eqn3}) in the appropriate
temperature range.

\section{A more general case}

In the last section we considered the case in the absence
of level degeneracy. In that case the relaxation time can be expressed in
terms of the level spacing distributions. However when some of the
energy levels are degenerate, we were in general not able to write 
Eq. (\ref{result}) in a simple form like Eq. (\ref{eqn3}) in terms
of the level spacing probability, $p_{(i)}(S)$. In addition if the degeneracy 
$g_X$ of the energy levels rises exponentially as the number of the 
energy level increases we can not neglect the $\eta_2$ term even if the 
temperature is much smaller than $\bigtriangleup E_{o}$. In this more
general case, the level degeneracies and the energetics will compete to
control the relaxation. Thus we expect the relaxation time $\tau$ to have 
a non-monotonic behavior as the temperature is changed.  In particular a
characteristic temperature $T_f$ is found which roughly separates the
relaxation regime which is controlled by energetics and the 
regime controlled by entropy.

\subsection{The characteristic temperature}

We now examine the special case of level degeneracy   which is approximately 
realized in protein spectra~\cite{zwanzig} where 
$g_{X}\sim \gamma^{X}$ and $\gamma$ is a constant greater than unity in this 
case local free energy $F_{i}$ equals $F_{i}=E_{i}-i\ln{\gamma}$. 
In this case it is easy to show, using Eq. (\ref{result}), that
\[\tau\ =\ \eta_1+\eta_2\ =\ \left( N + \sum_{i=1}^{N-1}\int_{0}^{+\infty}
p_{(i)}(S)\exp{\left[i\ln{\gamma}-\frac{S}{T}\right]}dS\right)
-\frac{\sum_{i=1}^{N}
Z_{i}^{2}\exp{\frac{F_{i}}{T}}}{Z_{0}}\ .\]
It is easy to show that the $\eta_2$ term as a whole scales as
$\exp{N(\ln{\gamma}-\frac{\bar{U}}{T}-\frac{E_{0}}{NT})}$
for temperatures $T>\frac{\bar{U}}{\ln{\gamma}}$, and 
the temperature at which this term becomes order unity is 
given by $T_{f} \sim \frac{\bar{U}+\frac{E_o}{N}}{ln{\gamma}}$
which is comparable with the average level spacing and becomes
independent of the ground state energy $E_o$ as $N$ approaches infinity. 
For $T > T_f$, $|\eta_2|$ is a rapidly increasing function of $N$ 
and at $T\sim T_f$ the two terms $\eta_1$ and $\eta_2$ become
comparable.

Our results show that the relaxation time of the model involves
competition between the two terms $\eta_1$ and $\eta_2$. The fact
that $|\eta_1|$ becomes comparable with $|\eta_2|$ at $T_{f}$
implies a change of behavior as temperature $T$ is swept across $T_f$.
To gain further intuition concerning the behavior of
the system at $T=T_f$, we examined the limit for which
all the $N$ levels in the quasi-continuous part of the spectrum 
are equidistant that $U$ is the energy difference between any two 
adjacent levels. Again let $\Delta E_o$ be the distance between the native 
state and the lowest energy of the quasi-continuous part of the spectrum. The 
equilibrium probabilities of each level will then become
\[ P_{0}=\frac{\exp{\frac{\Delta E_o}{T}}}{\exp{\frac{\Delta E_o}{T}}+
\frac{K^{N}-1}{K-1}}\]
and
\[ P_{i}=\frac{K^{i}}{\exp{\frac{\Delta E_o}{T}}+
\frac{K^{N}-1}{K-1}}\]
where $K=\exp(\ln{\gamma}-\frac{U}{T})=\exp(\ln{\gamma}(1-\frac{T_f}{T}))$.
Here we have defined $T_{f}=\frac{U}{\ln{\gamma}}$. Because $N$ is large, 
this expression for $K$ together with the expressions for $P_0$
and $P_i$ above shows that for $T < T_{f}$ the population of 
the energy levels will be given by $P_{0} \sim 1$ and $P_{i} \ll P_{0}$ 
for all $i\in\{1,N\}$. The equilibrium state of the system is then 
the ground state of our spectrum. For $T > T_{f}$, on the other hand,
$P_{i} \ll P_{i+1}$ for all $i\in\{0,N-1\}$, so that the equilibrium 
state will be shifted to the upper part of the 
quasi-continuous spectrum. Hence for $T < T_{f}$ or $T > T_{f}$ the 
equilibrium state of the system is relatively well defined.  
In the case when $T\sim T_{f}$ all the levels in the spectrum
become almost equally probable and hence large fluctuations can be  expected. 
>From the kinetic point of view, this means that at $T\sim T_{f}$ the
fluctuations lead to slow relaxation and thus large values of $\tau$. 
In this sense this behavior is similar to the well known ``critical
slowing down'' found for critical phenomena. From this argument it 
follows that for 
$T\sim T_{f}$ one should expect occurrence of a maximum in the 
relaxation time of a system. Finally we note 
that the characteristic temperature can also be obtained 
by the convergence criterion for the series of Eq. (\ref{mm2}), namely 
$\alpha_{0}R_{1}=1$.

We may conclude from our discussions that as soon as our system 
has energy levels with increasing  degeneracies, there exists a 
temperature scale $T_f$ 
which is determined by the average level spacing of the spectrum and the
degeneracy parameter $\gamma$. Close to $T_f$ the tendency of increasing 
entropy overcomes the tendency of relaxing into a state with lowest 
energy, and this leads to a sharp increase in the relaxation time 
of our system (see below for numerical calculations). 
Thus we expect a change in the behavior of the relaxation as temperature 
is varied by crossing $T_f$.  In this sense, the characteristic 
temperature $T_f$ can be regarded as the ``folding temperature''. 
This is indeed what observed in Ref. \cite{zwanzig} on situations similar
to that discussed in the last paragraph. Numerical results for 
this case will be presented in the next subsection.

For $T\ll T_{f}$, we can again neglect the $\eta_2$ term, and the 
relaxation time can be expressed as follows
\begin{equation}
\tau=t_{rel} \sim N + \sum_{i=1}^{N-1}\int_{0}^{+\infty} 
p_{(i)}(S)\exp{\left[i\ln{\gamma}-\frac{S}{T}\right]}dS\ \ \ .
\label{tau3}
\end{equation}
Similar to the discussion in the last section, it is clear that as the 
temperature increases from zero to $T_{f}$, $\tau$ becomes increasingly 
dependent on the higher order energy level spacing distributions. However when 
$T\sim 0$, $\tau$ is only determined by the nearest neighbor spacing 
distribution $p_{(1)}(S)$. Thus the relaxation at very low temperatures is
still generic in the sense of the discussion of last section,
namely $\tau$ is insensitive to the density of states $D(E)$.

For proteins in general, it is reasonable to assume that 
the number of levels is large, {\it i.e.} that $N\rightarrow \infty$. 
Then the characteristic temperature $T_f$ is of the order of
$\frac{\bar{U}}{ln \gamma}$. In order to examine higher temperatures, 
we first consider $T > T_f$ and the $\eta_1$
term as given by Eq. (\ref{tau3}). In this case the exponential in the 
integrand of Eq. (\ref{tau3}) is larger for larger 
values of the summation index $i$. 
Therefore for $T>T_{f}$, only higher order energy level spacing distributions
$p_{(i)}(S)$ play a substantial role.  However, as mentioned before, 
$p_{(i)}(S)$ approaches a Gaussian for large $i$ and hence the $\eta_1$ part
of the relaxation time again becomes generic for $T > T_f$. In addition, 
it is easy to see that the leading contribution 
in the $\eta_2$ term is due to the difference in energies 
between the highest and the lowest energy levels
for $T > T_f$ and is of the order of $\exp{N\frac{<S>}{T}}$. Since 
this also defines quite general properties of a given model, we expect 
weak model dependence for the $\eta_2$ term as well.
Hence we conclude that above the characteristic temperature, the relaxation
time is only weakly dependent on the detailed features of a given model.

The major model dependence is expected in the range 
$0\ll T\lesssim T_{f}$, while $T_{f}$ itself is determined by the 
degeneracy parameter $\gamma$ and average level spacing. 
The striking feature of the current case is that as $T$ reaches 
$T_{f}$ there is an increase in the dependence of $\tau$ on the higher 
order energy level spacing distribution:  at very low $T$ it is $p_{(1)}(S)$
which determines $\tau$, while above $T_f$ it is the higher $p_{(i)}(S)$
which is responsible. The important conclusion is that it is possible for 
different models which correspond to different density of states $D(E)$,
to have similar spacing distributions $p_{(1)}(S)$. If this is the case
the folding time is weakly sensitive to the details of a given 
model in the lower temperature range $T\ll \bar{U}$ for the non-degenerate 
models and, and $T\ll T_{f}$ for the degenerate models.

\subsection{Numerical results}

Although we have obtained all our results analytically, it is useful 
to obtain some numerical data as this gives considerable intuition about 
the relaxation kinetics of the model studied here.  For this purpose, we 
employ the model of Ref. \cite{zwanzig}.  In particular we assume 
that it has $N$ energy levels with equal nearest-neighbor spacings 
$U$ in its quasi-continuous part and $\Delta E_o$ below is the
ground state.  We assume $\Delta E_o \gg U$. 
The degeneracy of the quasi-continuous part is 
given by $\gamma^{i-1}$ where $i$ is the index of energy level. 
In the calculations we used $N=100,\ U=1$, and the energy gap between the
ground state and the bottom of the quasi-continuous part of the spectrum
$\Delta E_o=12$.

Then from our general result given by Eq. (\ref{result}), the relaxation 
time can easily be calculated and we obtain, for the whole temperature range,
\begin{equation}
\tau=N\frac{K^{N}-1}{K-1}-K\left(\frac{K^{N}-1}{K-1}\right)^{'}
-\sum_{i=1}^{N} \left(\frac{K^{N-i+1}-1}{K-1}\right)^{2} K^{i-1}
\frac{\alpha_{0}}{1+\alpha_{0}\frac{K^{N}-1}{K-1}}\ \ ,
\end{equation}
Here $K\equiv \gamma\exp{-\frac{U}{T}}$ and the prime means
differentiation with respect to $K$.  A simpler expression can be
obtained when the limit $N\rightarrow \infty$ is taken.

The relaxation time $\tau$ as a function of temperature for various 
degeneracy parameters $\gamma$ is shown in Fig. (\ref{tau}). We 
computed $\tau$ in two ways: either from a direct numerical inversion
of the matrix $(M-\alpha_0\delta M)$ of Eq. (\ref{inverse}) 
and then finding its trace, or by using the analytical form of 
Eq. (\ref{result}). Fig. (\ref{tau}) shows that these two 
methods give exactly the same results throughout the whole 
temperature range, justifying the mathematical procedure which led
to Eq. (\ref{result}). Several observations are in order.  First, when
the energy levels are non-degenerate, {\it i.e.} when
$\gamma =1$, there are no entropic effects in the model and the system
simply roles down the perfect funnel landscape in the relaxation 
process. In this case there is no characteristic temperature $T_f$ and
$\tau$ is completely determined by the energy level spacing 
distribution, as discussed before. Secondly, for cases with increasing 
level degeneracies, {\it i.e.} for $\gamma > 1$, the relaxation time
shows the expected maximum.  Also the position of the maximum
is exactly at the characteristic temperature $T_f$ (see below).
The behavior is consistent with that reported in Ref. \cite{zwanzig}.
Finally, the ``transition'' at $T_f$ becomes sharper as $\gamma$ is
increased. This is expected as it is similar to the situation that occurs
in a finite system where a thermal phase ``transition'' becomes 
sharper when the degree of freedom is increased.

In Fig. (\ref{tf}) the characteristic temperature as obtained from
$T_{f}\approx\frac{U}{\ln{\gamma}}+\frac{\bigtriangleup
E_{o}}{N\ln{\gamma}}$ is shown as a function of the degeneracy 
parameter $\gamma$. From this expression $T_f$ decreases monotonically 
as $\gamma$ is increased which must be true because higher degeneracies 
of energy spectrum leads to higher entropies involved. The data points 
in this figure were taken from the peak positions of Fig. (\ref{tau})
and are in good agreement with the theoretical definition.  Hence we 
conclude that $\tau$ takes maximum values at the characteristic
temperature $T_f$.

\subsection{A discussion on the effect of roughness}

So far our analysis is rigorous when the energy landscape is a perfect
funnel in the absence of roughness. Including arbitrary roughness will
make the problem essentially unsolvable analytically.  However under the
assumption that the roughness is small, our analysis can be extended to
estimate the effects of it. In this section we
will not attempt a rigorous treatment of the influence of roughness on 
the polymer dynamics. Rather, we will specify in what way the results of our 
theory will be modified if roughness is included. Our analysis follows the 
work of Leite and Onuchic\cite{Leite}.

The energy landscape roughness can be modeled\cite{Leite} by a distribution 
of states at a given value of the reaction coordinate $X$. The roughness is 
considered small if the width of the distribution is smaller than 
the average energy level spacing in the spectrum. Then, each energy
level $E(X)$ considered so far can be thought as being ``smeared'' out by
an energy probability distribution
$g(X,E)=\frac{1}{(2 \pi \delta E(X)^{2})^{1\over{2}}}
\exp{-\frac{(E-\bar{E}(X))^{2}}{2\delta E(X)^{2}}}$. Here
$\delta E(X)$ characterize the ``strength" of the roughness or the width 
of the energy band corresponding to a particular value of $X$. For this
rougness model, following Ref. \cite{Leite}, a
useful concept which arises is a coordinate-dependent phase 
transition\cite{Leite}. After the introduction of small roughness, 
the narrow band of states within $\delta E(X)$ can be considered to be a 
subsystem with its own dynamics. This consideration predicts\cite{Leite} 
that for an energy band with coordinate $X$ there is a critical temperature
$T_{c}(X)=\frac{\delta E(X)}{(2 \ln{\Omega (X)})^{1 \over{2}}}$ where
$\Omega (X)$ is a number of conformations corresponding to the 
the level at $X$. If $T \le T_{c}(X)$ for a particular $X$, the band at 
$\bar{E}(X)$ will behave in such a way that the dynamics inside this band is 
glass-like. This means that the system will tend to be frozen in a few low 
lying states of this band while relaxing inside it. This effect will have
an important influence on the relaxational dynamics. In the case of protein 
folding we model $\Omega (X) \sim \gamma^{X}$, and therefore
$T_{c}(X)=\frac{\delta E(X)}{(2 X \ln{\gamma})^{1 \over{2}}}$ where 
$X=0,1,2,...$. 

Although $\delta E$ may depend explicitely on $X$, let us first consider 
the limiting case when $\delta E$ is a constant over the spectrum. 
In this case for system temperature 
$T \le \frac{\delta E}{(2 X \ln{\gamma})^{1 \over{2}}}$ 
there are $I_f = (\frac{\delta E}{T})^{2} \frac{1}{2 \ln{ \gamma}}$ 
low lying energy bands with ``glass''-like dynamics. The lower the
temperature, the more ``frozen" bands would be in the system. As the 
global dynamics involves total number of energy states which is
proportional to $\sum_i^N\gamma^i=\frac{\gamma^{N+1}}{\gamma -1}$, 
a ``global'' phase transition temperature can be estimated, following 
Ref. \cite{Leite}, as
$T_{c}^{g} \sim \frac{\delta E}{(2 N \ln{\gamma})^{1 \over{2}}}$ 
where $N$ is the number of energy bands ($N=max\{X\}$).
It is important to compare this temperature scale with the folding temperature 
discussed previously, $T_{f} \sim \frac{\bar{U}}{ln{\gamma}}$. The number of 
``frozen" levels at the folding temperature is
$I_{f} \sim (\frac{\delta E}{\bar{U}})^{2}\frac{\ln{\gamma}}{2}
=T_c^g N^{1\over{2}}/T_{f}$.  This number can serve as a criterion for the 
definition of roughness ``strength". One can easily see that if the roughness 
$\delta E \lesssim \bar{U}$, then $I_{f} \sim 1$. We thus conclude that
only a few low lying energy bands with $X \sim 1$ are frozen at $T_f$. 
In this case the results derived from a perfect funnel should be applicable
for temperatures $T\sim T_{f}$, and the relaxation kinetics can be slowed 
down in the lower part of a spectrum near the ground state. 

The above discussion is for a constant $\delta E$ and more realistically there
is $X$ dependence of this quantity. For protein models it is natural to use
the measure of compactness, such as the total number of nearest contacts, to
specify the energy states. If $C_L$ is the maximum number of contacts for a
polymer with $L$ monomers, then we may define $C_L-X$ to give the total number 
of contacts at energy $E(X)$. It is reasonable to assume that for 
a structure with larger number of contacts, more interaction parameters are 
involved in computing the energy. In general for a given distribution of 
these parameters, the energy band width $\delta E$ is thus larger. For this
reason we expect a decreasing $\delta E(X)$ as $X$ is increased (although 
the level degeneracy $\gamma^X$ is increasing).  Hence the validity of the
perfect funnel results can be assessed by using the largest $\delta E(X)$
appropriate to the lower part of the spectrum and the discussion of the last
paragraph.

\section{Summary}

This work was motivated by a particular protein folding scenario based on
a funnel-like energy landscape. We studied the 
relaxational behavior when a system possesses a perfectly smooth
energy funnel. Our model is specified by a density of states $D(E)$ for
the quasi-continuous part, and a very low ground state level $E_o$. The
levels of the quasi-continuous part may have exponentially increasing 
degeneracies charactered by a degeneracy parameter $\gamma$. Because the
funnel is smooth, the relaxation is very simple and can be obtained by
using Metropolis rates. Indeed, the problem has been solved in closed form
for the whole temperature range in the aabsence of roughness. 

We found that the dependence of the relaxation time, $\tau$, on 
the energy level and the level spacing distributions of the models 
displays three main types of behavior depending on the temperature $T$.  
In the case where the energy levels are non--degenerate,
a general formula can be obtained relating $\tau$ to the level spacing
distribution.  Because $\tau$ is largely speaking determined by
the nearest level spacing distribution, which we have shown to be only
weakly dependent on $D(E)$, we conclude that the relaxation behavior 
can be said to exhibit ``universal'' features. In the degenerate case
which is more realistic for protein models, a characteristic temperature 
$T_f$ is found which separates the relaxation regimes dominated by energetics
or by entropic effects.  Hence in this sense $T_f$ can be thought of as
a ``folding temperature'' for the model studied here.  We found that
$\tau$ is weakly dependent on the specific density of states of a given model
if $T\ll T_{f}$ while becoming more and more model sensitive as $T$ approaches 
$T_{f}$. Again, at $T>T_{f}$ the system becomes almost insensitive to the 
detailed features of $D(E)$.  Across $T_{f}$ the relaxation time 
shows a maximum, indicating the ``folding transition'', in agreement with 
the earlier model of Zwanzig~\cite{zwanzig}. A possible experimental study 
of the relationship between the energy level distribution and folding 
kinetics as examined by our model would involve a systematic study of 
several proteins using a combination of thermodynamic techniques such as
calorimetry~\cite{griko} and folding assays.
 
The discussion presented here is only valid for a perfect funnel-like energy
landscape, {\it i.e.} that of a ``good'' protein sequence. The connection
to a particular model is through the density of states $D(E)$ which can
always be obtained numerically. If small amount of roughness is added via 
a finite width of the energy state distribution for a particular reaction 
coordinate $X$, the validity of the perfect funnel results can be examined
following the approach of the recently published work of Ref. \cite{Leite}.
In general, our theory correctly describe the kinetics of the model system 
including roughness in a tempearture range $T>N^{1\over{2}}T_{c}^{g}$:
below this temperature kinetics slows down by the roughness and 
approach glassy-like dynamics at $T \sim T_{c}^{g}$. As discussed in the last
section, $T_c^g$ is small for small amount of roughness (small $\delta E$).

Finally we comment that while our work was motivated by
the protein folding problem, the model is however only specified by the
density of states and relaxation in a perfect funnel with a 
low lying ground state. Thus the formula derived here are applicable to any 
other situations where a similar arrangement applies. On the other hand, 
as far as protein folding is concerned, the model studied here possesses 
many features of more realistic models.

\section*{Acknowledgments}

We thank Professor Martin Grant for a discussion concerning the master
equation.  One of us (HG) has benefited from discussions 
with Professor Minggao Gu concerning the level spacing statstics.
We gratefully acknowledge support by the Natural Sciences and
Engineering Research Council of Canada and le Fonds FCAR du Qu\'ebec
through centre and team grants.

\begin{figure}
\caption{A schematic plot of the perfect funnel--like energy landscape 
in reaction coordinate space. The energy spectrum has a low lying 
ground state with energy $E_o$, and a quasi-continuum part which is
specified by the levels $E_i$ with $i\in [1,N]$.
\label{funnel}
}
\end{figure}

\begin{figure}
\caption{
The logarithm of the relaxation time $\tau$ as a function of temperature 
$T$ for various degeneracy parameters $\gamma$ as indicated by the numbers
near the curves. The unit of $T$ is the level spacing $U$, and arbitrary
unit is set for $\tau$. The solid squares 
were computed from a direct numerical inversion of the matrix 
$(M-\alpha_0\delta M)$ of Eq. (\ref{inverse}) followed by 
a calculation of the trace of $(M-\alpha_0\delta M)^{-1}$. 
The solid lines were calculated using the analytical form of
Eq. (\ref{result}). Clearly these two methods give exactly the same 
answers throughout the whole temperature range. 
\label{tau}
}
\end{figure}

\begin{figure}
\caption{
The characteristic temperature $T_f$ as a function of the level degeneracy
parameter $\gamma$.  Unit of $T_f$ is the level spacing $U$.  The solid 
line is obtained from $T_{f}\approx\frac{U}{\ln{\gamma}}+\frac{\bigtriangleup
E_{o}}{N\ln{\gamma}}$. The solid squares in this figure correspond 
to the temperatures at which the relaxation time achieves its 
maximum, {\it i.e.} the peak positions of Fig. (\ref{tau}).
These two prescriptions give nearly the same values for $T_f$.
\label{tf}
}
\end{figure}


\begin{references}

\bibitem{creighton}
See, for example, articles in {\it Protein Folding}, Ed. T.E. Creighton, (W.H.
Freeman and Company, New York, 1992).

\bibitem{bryngelson}
J.D. Bryngelson and P. Wolynes, J. Phys. Chem. {\bf 93}, 6902 (1989).

\bibitem{go}
N. Go and H. Abe, Biopolymers {\bf 20}, 991 (1981).

\bibitem{montal}
Peter E. Leopold, Mauricio Montal, and Jos\'e Nelson Onuchic,
Proc. Natl. Acad. Sci. USA {\bf 89}, 8721 (1992).

\bibitem{wolynes}
P.G. Wolynes, J.N. Onuchic, D. Thirumalai, Science, {\bf 267}, 1619 (1995).

\bibitem{onuchic}
J.N. Onuchic, P.G. Wolynes, Z. Luthey-Schulten and N.D. Socci, Proc. Natl.
Acad. Sci. USA, {\bf 92}, 3626 (1995).

\bibitem{zwanzig}
R. Zwanzig, Proc. Natl. Acad. Sci. USA {\bf 92}, 9801(1995).

\bibitem{bryngelson1}
J.D. Bryngelson and P.G. Wolynes, Proc. Natl. Acad. Sci. USA {\bf 84}, 7524
(1987).

\bibitem{honeycutt}
J.D. Honeycutt and D. Thirumalai, Proc. Natl. Acad. Sci. USA {\bf 87},
3526 (1990); Biopolymers {\bf 32}, 695 (1992);
Z. Guo, Thirumalai, and J.D. Honeycutt, J. Chem. Phys. {\bf 97}, 525
(1992).

\bibitem{shakhnovich2}
E.I. Shakhnovich and A.M. Gutin, Nature {\bf 346}, 773 (1990).

\bibitem{maxim1}
The details of the calculation will be presented elsewhere.
Maxim Skorobogatyy, unpublished.

\bibitem{mehta}
M.L. Mehta, {\it Random matrices and the statistical theory of energy levels},
(Academic Press, New York, 1967).

\bibitem{abkevich}
V.I. Abkevich, A.M. Gutin and E.I. Shakhnovich, J. Chem. Phys. {\bf 101},
6052 (1994).

\bibitem{pandey}
A. Pandey, Annals of Physics {\bf 119}, 170 (1979)

\bibitem{griko}
Y. V. Griko, E. Freire, G. Privalov, H. van Dael and P.L. Privalov,
J. Mol. Biol. {\bf 252}, 447 (1995)

\bibitem{Leite}
Vitor B. P. Leite and Jos\'e N. Onuchic, J. Phys. Chem. 100, 7680 (1996)

\end{references}
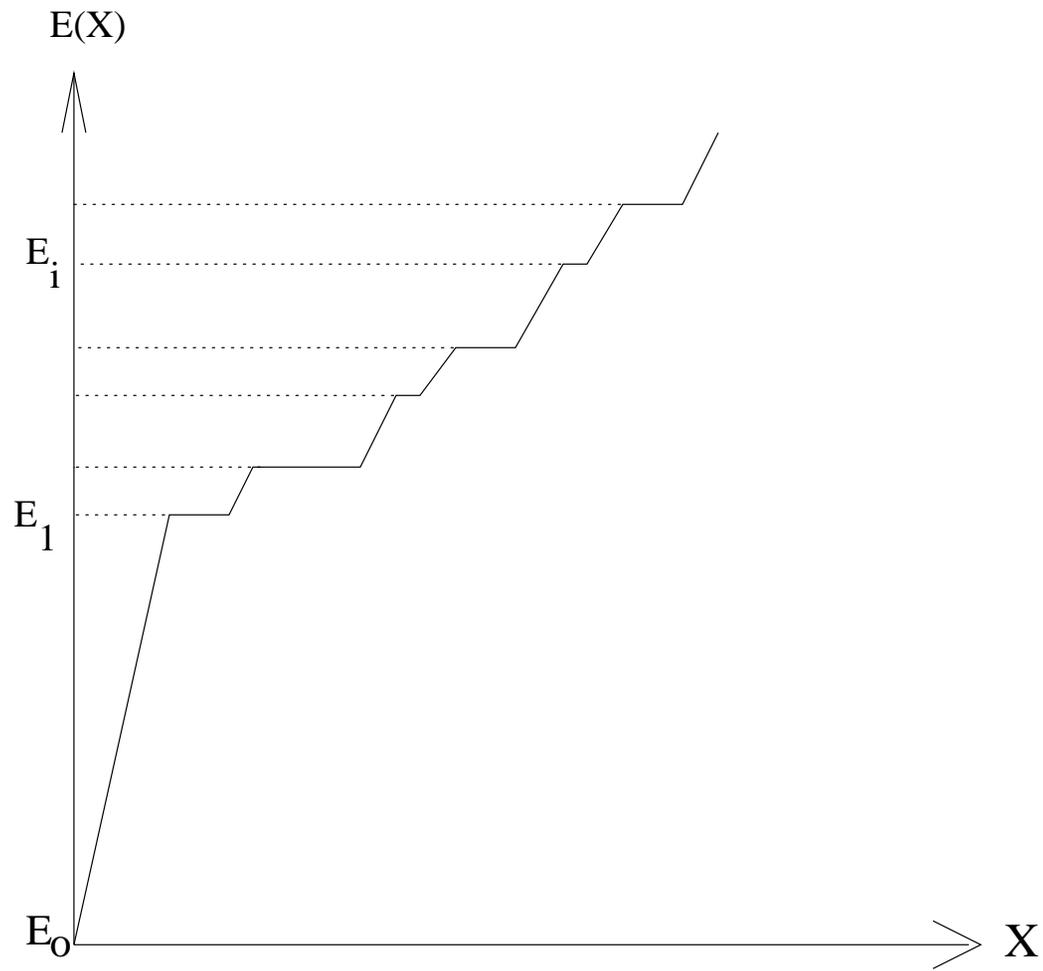
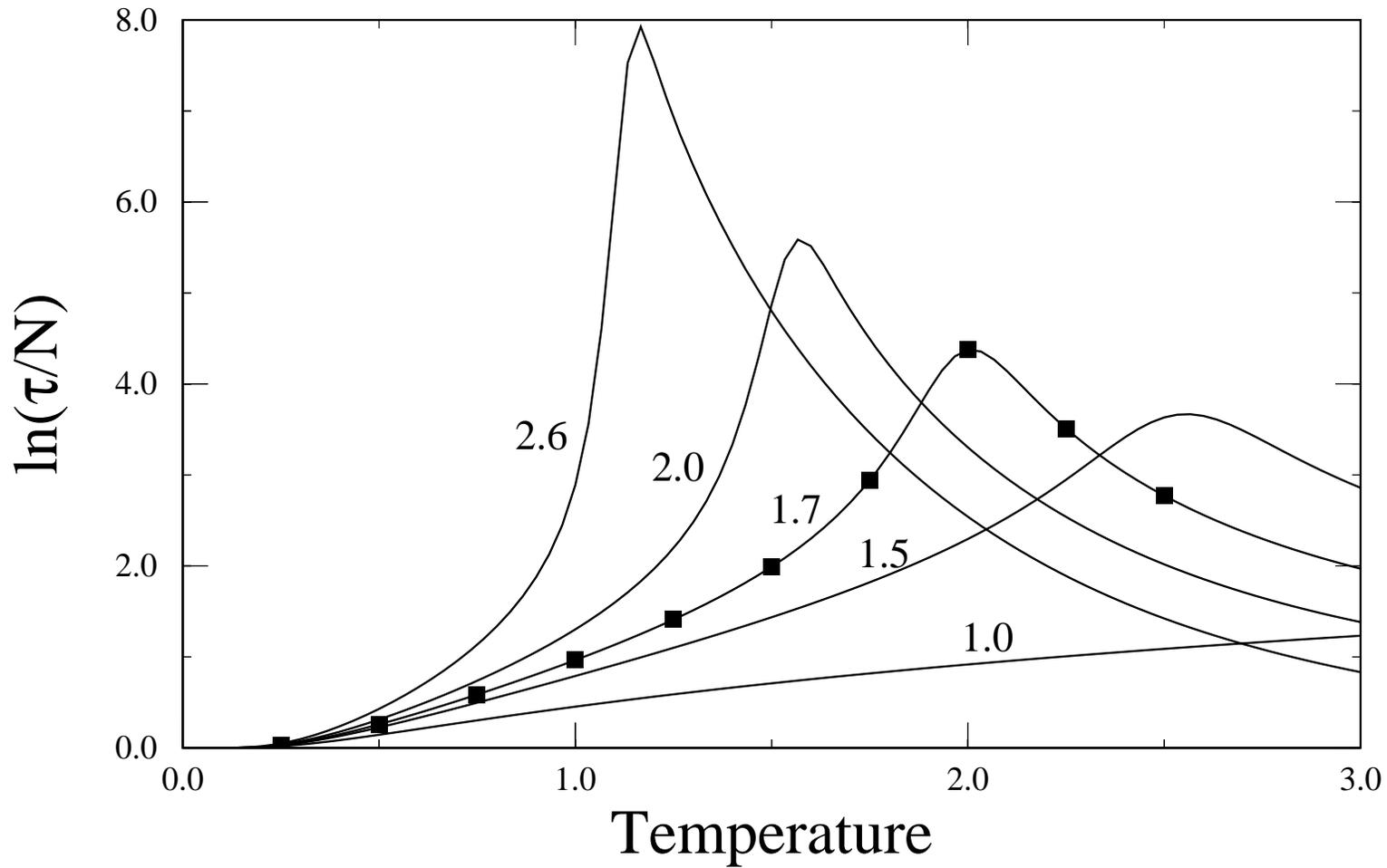
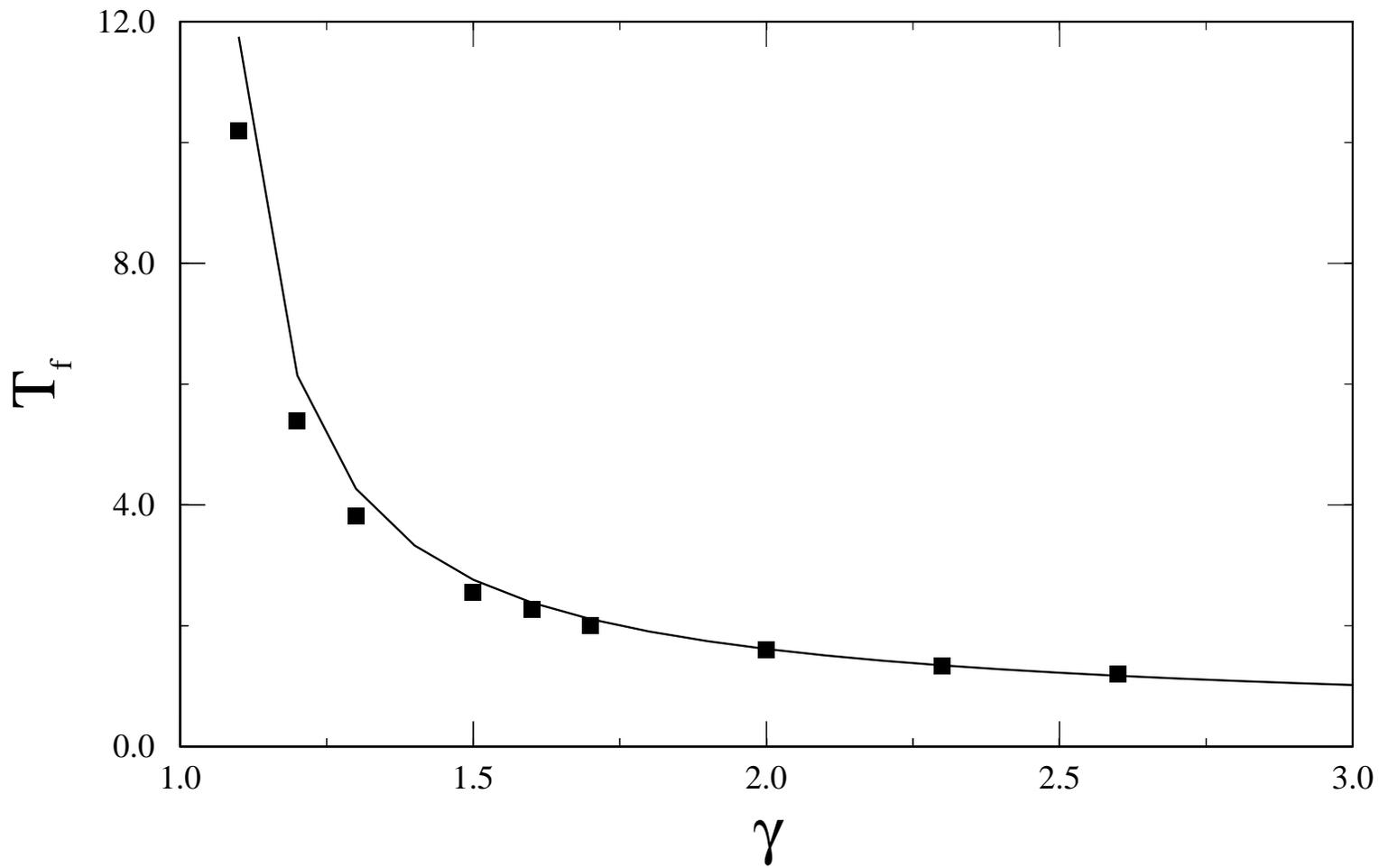
\end{document}